\documentstyle{ptptex}
\newcommand{\beq}{\begin{equation}}

\newcommand{\eeq}{\end{equation}}

\newcommand{\beqa}{\begin{eqnarray}}
\newcommand{\eeqa}{\end{eqnarray}}
\newcommand{\ba}{\begin{array}}
\newcommand{\ea}{\end{array}}

\newcommand{\re}{{(E-x^N)}}

\def\fnote#1{\footnote}

\newcommand{\be}{\begin{equation}}

\newcommand{\ee}{\end{equation}}

\newcommand{\s}{\sigma}
\newcommand{\ii}{{\rm i}}

\title{%        %You can use \\ for explicit line-break
 On the semiclassical  expansion for 1-dim $x^N$ potentials}

\author{%       %Use \sc for the family name
Marko Robnik\footnote{e--mail: robnik@uni-mb.si}
and
Valery Romanovski\footnote{e--mail: valery.romanovsky@uni-mb.si}
}

\inst{%         %
Center for Applied Mathematics and Theoretical Physics,\\
University of Maribor, Krekova 2, SI-2000 Maribor, Slovenia
}

\abst{%       %this abstract is neglected when [addenda] or [errata]
We obtain a simple formula for  the  semiclassical series
for potentials $V(x)=x^N$ ($N$ even) and
derive almost  explicit formulae for the WKB
approximation of  the
energy eigenvalues of such potentials.
}

\begin{document}

\maketitle
In the present paper we  study  the structure of the
WKB series  for the polynomial potential
$V(x)=x^N$ ($N$ even). In particular, we obtain relatively simple
recurrence formula of the coefficients  $\s'_k$ of the semiclassical
approximation and of  the WKB terms for the
energy eigenvalues. We follow our aproach as expounded in 
ref. \cite{rf:RobRom1999}.

We consider the two-turning point eigenvalue problem for the
one-dimensional
 Schr\"o\-dinger equation
\be   \label{sch}
 [- {\hbar ^2} \frac {{\rm d}^2}{{\rm  d} x^2}+V(x)]\psi (x)=E\psi(x).
\ee

\noindent
We can always write the wavefunction as
\be
\psi(x) =\exp \left\{\frac 1\hbar \sigma (x)\right\}
\ee
where the phase $\sigma (x)$ is a complex function that satisfies the
differential equation
\be \label{w1}
\sigma '^2(x)+ \hbar  \sigma '' (x)=
(V(x)-E)\stackrel{def}{=} Q(x).
\ee
The WKB expansion for the phase is
\be \label{w2}
\sigma(x)=\sum_{k=0}^{\infty}
  \hbar^k \sigma_k (x).
\ee
Substituting (\ref{w2}) into (\ref{w1}) and comparing
like powers of  $\hbar$ gives the recursion relation
\be \label{w3}
\s _0'^2 = Q(x),\ \ \ \ \ \sum_{k=0}^n \s' _k \s ' _{n-k}+\s
_{n-1}''=0.
\ee

We assume that the potential $V(x)$ is single valued
analytic function which is real on the real axis, that
$V(\pm\infty)=\infty,$ that $V(x)$
has a unique minimum on the real axis, that $V(x)$ rises
monotonically
on both sides of the minimum and for $k>1$ the integrals
\be \label{kint}
   \int_{-\infty}^a |\s'_k(x) | dx,\  \int_{b}^{+\infty} |\s'_k(x) | dx
\ee
are convergent, when $a<x_1, b>x_2$, where  $x_1<x_2$ are the turning
points.
Then the following
asymptotic expansion takes place \cite{rf:Fedoryuk1983}:
\be \label{gin}
\frac 1{2\ii}\oint_\gamma \sum_{k=0}^{\infty}
\hbar^k  \sigma'_k(x)  dx  = \pi n \hbar,
\ee
where $n \ge 0$ is  an integer number
and $\gamma $ is a contour surrounding the turning points on the real
axis (for the first time this formula appears, apparently, in
\citen{rf:Dunham1932}).
This relation is an equation with respect to $E$ and using it one can
find the asymptotic of the eigenvalues $E_n(\hbar).$ In some
cases the series (\ref{gin}) can be summed exactly
(see e.g.  
\citen{rf:Bender1977,rf:RobnikSalasnich1997a,rf:RobnikSalasnich1997b,rf:RomanovskiRobnik1999,rf:SalasnichSattin1997}).

The zero-order term of the expansion (\ref{kint}) is given by
\be
\frac 1{2 \ii} \oint _\gamma { d} \sigma_0 = \int { d} x
\sqrt{E-V(x)},
\ee
the first odd term is
\be
 \frac \hbar{2 \ii}
  \oint _\gamma{ d } \sigma_1 =-\frac{\pi \hbar}2
\ee
and to find the higher order terms we need to compute
 the functions $\s'_k$ using the recursion relation (\ref{w3}).

We now consider the potentials $V(x)=x^N,$ where $N$ is even positive
integer.
Let us show  that for such
potentials the coefficients $\s'_k$\ $(k \ge 1)$ of the WKB expansion
have the form
\be \label{sxn}
\s'_k=-\frac {(-\ii)^{3k-1} x^{-k+N}}{(E-x^N)^{\frac{3k-1}2}}
\sum_{j=0}^{k-1} A_{k-j-1,j} E^{k-j-1} x^{jN},
\ee
where  we choose $
\sqrt{E-x^N}=\ii \sqrt{x^N-E}$, and the coefficients $A_{k-j-1,j}$ of
the monomials $ E^{k-j-1} x^{jN}$ are computed according to the
recurrence
formula
\begin{eqnarray} \label{fa}
\lefteqn{A_{s,l}=\frac 12 \sum_{i=0}^s \sum_{j=0}^{l-1}
 A_{i,j} A_{s-i, l-1-j}+}\\
& &
  \frac{l(2+N)+(2+3N) s-N}4 A_{s,l-1}+\frac {(N-1) l +N-s} 2
A_{s-1,l}\nonumber.
\end{eqnarray}

We   prove formula (\ref{fa}) by induction. We have
\be
  \s'_0=\ii \sqrt {E-x^N},\  \s'_1=\frac{Nx^{N-1}}{4\re},
\ee
it means that for $k=1$ formula (\ref{fa}) takes place,
and also $A_{0,0}= N/4$.
Let us suppose that it holds  for $k<m.$ Then for $k=m$
in the case $0<u<m$ we have
\be \label{f1}
  \s'_u \s'_{m-u}=\frac{(-\ii)^{3m-2}x^{-m+2N}}{\re^{\frac{3m-2}2}}
  \sum_{j=0}^{u-1}A_{u-j-1,j}E^{u-j-1}x^{jN} \sum_{r=0}^{m-u-1}
  A_{m-u-r-1,r}E^{m-u-r-1}x^{rN}.
\ee
Note that for a fixed $i,j$ the coefficient $A_{i,j}$ is a coefficient
 in the expresion (\ref{sxn}) for $\s'_t$ if and only if $i+j=t-1$.
Then from (\ref{f1}) we get that the contribution into
\be
\frac{(-\ii)^{3m-2}x^{-m+N}}{\re^{\frac{3m-2}2}}
 A_{s,l} E^{s}x^{lN}
\ee
$(s+l=m-1)$ arises from
\be \label{f2}
\sum_{i+j=u-1}\frac{(-\ii)^{3m-2}x^{-m+2N}}{\re^{\frac{3m-2}2}}
A_{i,j} E^ix^{jN} A_{s-i,l-j-1} E^{s-i}x^{(l-j-1)N}=
\ee
$$
\frac{(-\ii)^{3m-2}x^{-m+N}}{\re^{\frac{3m-2}2}}
\sum_{i+j=u-1} A_{i,j}  A_{s-i,l-j-1} E^{s}x^{lN},
$$
here $s-i+l-j-1=m-u-1$ and, therefore, $A_{s-1,l-j-1}$ is a coefficient in
$\s'_{m-u}.$ Next,
\be     \label{f3}
\s''_{m-1}=-
\frac{(-\ii)^{3m-4}x^{-m+N}}{\re^{\frac{3m-2}2}}
\sum_{j=0}^{m-2}[(1-m+N+jN)A_{m-j-2,j} E^{m-j-1}x^{jN}+
\ee
$$
(-1-3N+m+\frac 32 mN-jN)A_{m-j-2,j} E^{m-j-2}x^{(j+1)N}].
$$
Hence, from (\ref{w3}), (\ref{f2}), (\ref{f3}) we conclude,
that $A_{s,l}$ is computed according (\ref{fa}) and
$\s'_m$ has the form (\ref{sxn}).

Consider the formal series
\be
   A(z_1,z_2)=\sum_{i+j\ge 0} A_{i,j}z_1^iz_2^j,
\ee
where $A_{i,j}$ are numbers defined by the recurrence formula
(\ref{fa}). Using the properties of generating functions
(see, e.g. \cite{rf:Graham1994}) we obtain from
the recurrence relation (\ref{fa})
the following differential equation for the generating function
$A(z_1,z_2)$
\begin{eqnarray} \label{urav} \nonumber
A=\frac 12 z_2 A^2 +\frac{2+N}4 z_2(z_2 A )'_{z_2}+\frac {2+3N}4 z_1
z_2  A'_{z_1}-\\
\frac N4 z_2 A+\frac {N-1}2 z_1 z_2 A'_{z_2}-\frac 12
z_1(z_1A)'_{z_1}+\frac N2 z_1 A+\frac N4.
\end{eqnarray}

We cannot  solve this equation in the general case, but we
can use it in order to find the coefficients of the form $A_{i,0}$.
Indeed, if we put   $z_2\equiv 0$, then equation (\ref{urav}) has the form
\be
   z^2 A'(z)-(z (N-1)-2) A(z)- N/2=0,
\ee
where we set $z=z_1.$
The general solution of this equation is
\be
 A(z)=
     C e^{\frac{2}{z} - \left( 1 - N \right) \,\log (z)}\,
         + 2^{-1 - N}\,e^{\frac{2}{z}}\,N\,
        z^{ N-1}\,\Gamma(N,\frac{2}{z}),
\ee
where $\Gamma(n,z)=\int_z^\infty t^{n-1}e^{-t}dt.$
However to satisfy the initial condition
$A(0)=N/4$ we have to put $C=0.$ Therefore the coefficients
$A_{k,0}$ are the ones  of Taylor expansion of the
function
\be
 2^{-1 - N}\, e^{{\frac{2}{z}}}\,N\,
  {{{{z}}}^{ N-1}}\,\Gamma(N,\frac{2}{z}).
\ee
Integrating we get for integers $N>0$
\be
 2^{-1 - N}\, e^{{\frac{2}{z}}}\,N\,
  {{{{z}}}^{ N-1}}\,\Gamma(N,\frac{2}{z})=
  \sum_{k=1}^{N}\frac {N!}{2^{N-k+2}(k-1)!}z^{N-k}.
\ee
Therefore
\be
   A_{s,0}=\frac{N!}{2^{s+2}(N-s-1)!},
\ee
and this is in  agreement with $A_{0,0}=N/4$.
Unfortunately, we are not able to  find the solution of the general equation
(\ref{urav}) and, hence, cannot get similar expressions for other
coefficients $A_{s,l}.$

However using  relations (\ref{sxn}), (\ref{fa})
 we can  obtain the following formula for the
coefficients of the  WKB expansion
\begin{eqnarray}\nonumber
\oint_\gamma  \s'_{2k} dx=\frac { \ii 2^{3k+1}\sqrt \pi}{(6k-3)!! N}
E^{\frac 12+\frac 1N-\frac{2k}N-k}
\frac {\Gamma(\frac{1-2k}N+1)}{\Gamma(\frac {3-2k}2
+\frac {1-2k}N)} (A_{2k-1,0}\prod_{s=1}^{2k-1} (\frac{3-2k}2+\\  
\frac{1-2k}N-s)+\sum_{i=1}^{2k-1}
 A_{2k-i-1,i} \prod_{s=1}^i (s+\frac{1-2k}N)
\prod_{s=1}^{2k-i-1}(\frac{3-2k}2+\frac{1-2k}N-s ), \nonumber\\
\label{23}
\end{eqnarray}
where $k\ge 1$ and $ A_{2k-i-1,i}$ are computed according to (\ref{fa})
and
\be
\oint_\gamma \s'_0 dx=\frac{2\ii E^{\frac 12+\frac 1N} \sqrt \pi \Gamma
(1+\frac 1N)}{\Gamma (\frac 32+\frac 1N)}.
\ee

>From (\ref{sxn}) we conclude that the integrals (\ref{kint})
are convergent for all $k>1$ and according to the above mentioned
statement from \cite{rf:Fedoryuk1983}  formulae (\ref{gin}) and
(\ref{23})  can be used to find the asymptotic approximation
of the eigenvalues $E_n(\hbar)$. We should emphasize that
quite generally the odd-order terms (except first order for
$\sigma'_1$) yield integrals that vanish exactly, because,
as it follows from the results of Fr\"oman \cite{rf:Froeman1966},
$\sigma'_{2k+1}$ are total derivatives for all $k>0$.

To summarize,  in this paper
we derive relatively simple  explicit formula for the WKB terms for the
energy eigenvalues of the homogeneous power law potentials
$V(x) = x^N$, where $N$ is even.
 In computing by means
of this formula we manipulate only with {\em numbers} and do not need to
collect similar terms of a polynomial, which we must do otherwise
when we use just the recursion formula (\ref{w3}).
As it is known the operation of collection of similar terms
is very laborious even for Computer Algebra systems.
So,  application of the
obtained formulae considerably simplifies calculations, especially if we
need to compute high order terms.

\section*{Acknowledgements}
This project was  supported by the Ministry of Science and Technology of
the Republic of Slovenia and by the  Rector's Fund of the University of
Maribor.
VR acknowledges the support of the work by the grant of the Ministry of
Science
and Technology of the Republic of Slovenia and  the Abdus Salam   ICTP
(Trieste) Joint Programme and also the support of  the Foundation of
Fundamental Research of the
Republic of Belarus.

\end{document}